\documentclass[aps,prb,showpacs,twocolumn,superscriptaddress]{revtex4-1}
\usepackage{amsmath}
\usepackage{amssymb}
\usepackage{graphicx}
\begin{document}
\title{Microscopic model for the ferroelectric field effect in oxide heterostructures}
\author{Shuai Dong}
\affiliation{Department of Physics, Southeast University, Nanjing 211189, China}
\affiliation{National Laboratory of Solid State Microstructures, Nanjing University, Nanjing 210093, China}
\author{Xiaotian Zhang}
\affiliation{Department of Physics and Astronomy, University of Tennessee, Knoxville, Tennessee 37996, USA}
\affiliation{Materials Science and Technology Division, Oak Ridge National Laboratory, Oak Ridge, Tennessee 32831, USA}
\author{Rong Yu}
\affiliation{Department of Physics and Astronomy, Rice University, Houston, Texas 77005, USA}
\author{J.-M. Liu}
\affiliation{National Laboratory of Solid State Microstructures, Nanjing University, Nanjing 210093, China}
\affiliation{International Center for Materials Physics, Chinese Academy of Sciences, Shenyang 110016, China}
\author{Elbio Dagotto}
\affiliation{Department of Physics and Astronomy, University of Tennessee, Knoxville, Tennessee 37996, USA}
\affiliation{Materials Science and Technology Division, Oak Ridge National Laboratory, Oak Ridge, Tennessee 32831, USA}
\date{\today}

\begin{abstract}
A microscopic model Hamiltonian for the ferroelectric field effect is introduced
for the study of oxide heterostructures with ferroelectric components. The long-range
Coulomb interaction is incorporated as an electrostatic potential, solved self-consistently
together with the charge distribution. A generic double-exchange system is used as the
conducting channel, epitaxially attached to the ferroelectric gate. The observed ferroelectric
screening effect, namely the charge accumulation/depletion near the interface, is shown to drive
interfacial phase transitions that give rise to robust magnetoelectric responses and bipolar
resistive switching, in qualitative agreement with previous density functional theory calculations.
The model can be easily adapted to other materials by modifying
the Hamiltonian of the conducting channel, and it is useful in simulating ferroelectric
field effect devices particularly those involving strongly correlated electronic components where
\textit{ab-initio} techniques are difficult to apply.
\end{abstract}
\maketitle

\section{Introduction}
The research area known as oxide heterostructures
continues attracting considerable attention of the condensed matter community
due to the rich physical properties of its constituents,
often involving strongly correlated electronic materials, and also for their broad potential in device
applications.\cite{Dagotto:Sci07,Takagi:Sci,Mannhart:Sci,Hammerl:Sci} Among these heterostructures, those involving ferroelectric (FE) and magnetic,
or multiferroic, components are particularly interesting since they could be used in the next
generation of transistors and nonvolatile memories.\cite{Ramesh:Nm,Bibes:Ap,Nan:Jap} From the applications perspective,
the FE/magnetic heterostructures could become even  superior to the currently available bulk multiferroics with regards
to their magnetoelectric performance.\cite{Ramesh:Nm,Bibes:Ap,Nan:Jap} In these heterostructures, it is easier to obtain
large FE polarizations and a robust magnetization, and the manifestations of the magnetoelectric coupling can be fairly
diverse. For example, an exchange bias effect that can be controlled with
electric fields  has been recently reported
in La$_{0.7}$Sr$_{0.3}$MnO$_3$/BiFeO$_3$ \cite{Wu:Nm10,Yu:Prl} and the associated physical
mechanism that produces this interesting behavior
is being actively discussed.\cite{Dong:Prl2,Belashchenko:Prl,Livesey:Prb,Okamoto:Prb10}

Even without the magnetic coupling across the interface,
interfacial magnetoelectric effects still generally exist
in these heterostructures. A mechanism contributing to these effects involves
the possibility of lattice distortions, since
the oxides magnetic or FE properties are often
sensitive to strain.\cite{Zheng:Jap,Zheng:Prb,Guo:Apl}
An additional contribution is the carrier-mediated field effect,\cite{Rondinelli:Nn,Duan:Prl}
especially crucial in ultrathin film heterostructures. The FE field effect not only generates
magnetoelectricity, but also gives rise to
a bipolar resistive switching.\cite{Ahn:Sci,Mathews:Sci,Hoffman:Am,Molegraaf:Am,Vaz:Prl,Vaz:Apl,Hong:Apl,Zhao:Apl,Thiele:Apl,Chaudhuri:Apl,Watanabe:Apl,Kuffer:Nm,Zayas:Prl}

\begin{figure}
\centering
\includegraphics[width=0.4\textwidth]{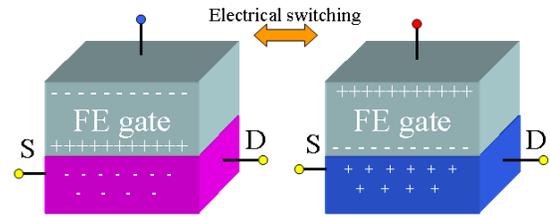}
\caption{(Color online) Sketch of a FE-FET heterostructure (S indicates the source and D the drain).
The physical properties of the channel can be switched on and off by the FE polarization of the gate.}
\label{fet}
\end{figure}

A heterostructure FE field-effect transistor (FE-FET) is basically composed of a FE oxide film and a thin metallic
or semiconducting oxide film, as sketched in Fig.~\ref{fet}, similarly to traditional FETs used in the semiconductor
industry. In those standard FET devices, the conductivity of the semiconducting
channel can be switched on and off
by tuning the gate voltage. The FE-FETs can provide similar functions by switching
the direction of the polarization of the FE gate.
Moreover, this switching, at least ideally, can be non-volatile due to the remnant FE polarization.\cite{Ahn:Sci,Hoffman:Am}
Furthermore, due to the strongly correlated character of the electronic component
in several oxides, the above mentioned switching in FE-FET is not limited
to the conductivity, but it may also influence on
other physical quantities as well, such as the magnetization, orbital order,
elastic distortions, etc. Therefore, compared with traditional semiconductor FETs,
the physics in FE-FETs can be richer,
and potentially additional functionalities can be expected.

Although the FE-FETs have been experimentally studied for several years, only recently theoretical investigations have
been focused on this topic.\cite{Rondinelli:Nn,Duan:Prl,Burton:Prb,Burton:Prl,Stengel:Prl,Bristowe:cm}
These recent theoretical studies have been based
on the density functional theory (DFT). In fact,
studies using model Hamiltonians including strongly correlated electronic
effects, beyond the reach of DFT,
and focusing on the basic aspects of the FE field effect in these oxide
heterostructures are rare. An important technical problem in this context is how to take into account
the contribution from the FE polarization on  the physics of the
microscopic model Hamiltonian representing the other components.
In recent efforts by some of us, the FE polarization was modeled as an interfacial potential at the first layer
of the conducting channel,\cite{Calderon:Prb11} but this approximation must be refined to address
the subtle energy balances between competing tendencies near the interface. Thus, for all these reasons
in this manuscript the FE-FET structures
will be revisited using model Hamiltonian
techniques and applying new approximations to handle this problem. Our effort has the main merit of
paving the way for the use of models for the study of FE-FET systems where one of the components
has a strongly correlated electronic character that is difficult to study via \textit{ab-initio} methods.

\section{Model and method}
\subsection{Model Hamiltonian}
As discussed in the Introduction, in this manuscript the FE field effect will be studied from
the model Hamiltonian perspective. More specifically, here the standard two-orbital (2O) double-exchange (DE) model
will be used for the metallic component of the heterostructure. This 2O DE model is well known to be successful in
modeling the perovskite manganites,\cite{Dagotto:Prp,Dagotto:Bok,Dagotto:Njp} which are materials  often used in FE-FET devices.
Furthermore, previous model Hamiltonian studies have already confirmed that the 2O DE model, with some simple modifications,
is still a proper model to use
for manganite layers when they are in the geometry of a heterostructure.\cite{Dong:Prb08.3,Brey:Prb,Yu:Prb09,Calderon:Prb08,Nanda:Prb,Calderon:Prb11}
In addition, since the DE mechanism provides a generic framework to describe the motion of electrons
in several magnetic systems, the approach followed here, with minor modifications, could potentially be adapted to other oxides beyond the manganites.

As a widely accepted simplification, the limit of an infinite Hund coupling will be adopted in the DE model studied here. Then, more specifically
the model Hamiltonian of the metallic channel reads as:
\begin{eqnarray}
\nonumber H&=&-\sum_{<ij>}^{\gamma\gamma'}t^{\vec{r}}_{\gamma\gamma'}(\Omega_{ij}c_{i\gamma}^{\dagger}c_{j\gamma'}+H.c.)+\sum_{i}V_{i}n_{i}\\
&&+\sum_{<ij>}J_{\rm AF}\vec{S}_{i}\cdot\vec{S}_{j}.
\label{H}
\end{eqnarray}

In this expression the first term is the standard DE interaction.
The operator $c_{i\gamma}$ ($c_{i\gamma}^{\dag}$) annihilates
(creates) an electron at the orbital $\gamma$ of the  $e_{\rm g}$ band and
at the lattice site $i$, with its spin perfectly parallel
to the localized $t_{\rm 2g}$ spin $\vec{S}_{i}$. The indexes
$i$ and $j$ represent nearest-neighbor (NN) lattice sites. The Berry phase factor $\Omega_{ij}$, generated by the infinite Hund coupling limit adopted here,
equals $\cos(\theta_{i}/2)\cos(\theta_{j}/2)+\sin(\theta_{i}/2)\sin(\theta_{j}/2)\exp[-i(\phi_{i}-\phi_{j})]$, where $\theta$ and $\phi$ are the polar and
azimuthal angles defining the direction of the $t_{\rm 2g}$ spins, respectively. When a ferromagnetic (FM) $t_{\rm 2g}$ background is used, then $\Omega=1$.
The labels $\gamma$ and $\gamma'$ denote the two Mn $e_{\rm g}$-orbitals $a$ ($|x^2-y^2>$) and $b$ ($|3z^2-r^2>$).
The NN hopping direction is denoted by $\vec{r}$. The DE hopping depends
on the direction in which the hopping occurs, and it is orbital-dependent as well. The
actual hopping amplitudes are:
\begin{eqnarray}
\nonumber t^x&=&\left(
\begin{array}{cc}
t^x_{aa} &  t^x_{ab} \\
t^x_{ba} &  t^x_{bb}
\end{array}
\right)=\frac{t_0}{4}\left(
\begin{array}{cc}
3 &  -\sqrt{3} \\
-\sqrt{3} &  1
\end{array}
\right)\\
\nonumber t^y&=&\left(
\begin{array}{cc}
t^y_{aa} &  t^y_{ab} \\
t^y_{ba} &  t^y_{bb}
\end{array}
\right)=\frac{t_0}{4}\left(
\begin{array}{cc}
3 &  \sqrt{3} \\
\sqrt{3} &  1
\end{array}
\right)\\
t^z&=&\left(
\begin{array}{cc}
t^z_{aa} &  t^z_{ab} \\
t^z_{ba} &  t^z_{bb}
\end{array}
\right)=t_0\left(
\begin{array}{cc}
0 &  0 \\
0 &  1
\end{array}
\right),
\label{Eq.hopping}
\end{eqnarray}
\noindent where $t_0$ is the DE hopping amplitude scale. In the rest of this publication, $t_0$ is considered the unit of energy. Its real value is approximately $0.5$~eV in wide-bandwidth
manganites such as La$_{0.7}$Sr$_{0.3}$MnO$_3$ (LSMO).\cite{Dagotto:Bok,Dagotto:Prp}

The second term in the Hamiltonian is the on-site potential energy: $V_i$ is the actual
potential at each site and $n_i$ is the $e_g$ electronic density operator at the same site.
The last term is the Heisenberg-type antiferromagnetic (AFM) superexchange (SE) interaction
between the localized NN $t_{\rm 2g}$ spins. Its actual typical strength
is about $10\%$ that of $t_0$.\cite{Dagotto:Prp,Dagotto:Bok}

\subsection{Self-consistent calculations}
In the actual calculations described in this publication, a cuboid lattice ($L_x\times L_y\times L_z$, $L_x$=$L_y$=$4$, $L_z$=$12$) will be used,
with open boundary conditions (OBCs) along the $z$-axis to avoid having two interfaces.\cite{Yu:Prb09,Calderon:Prb11}
Twisted boundary conditions (TBCs) are adopted in the $x$-$y$ plane to reduce finite size effects, via a $6\times6$ $k$-mesh.

The FE gate will be here modeled as a surface charge ($Q$ per site, in units of the elementary charge $e$, and located at $z$=$0$)
coupled to the first channel layer ($z$=$1$). This approximation has been successfully confirmed in previous DFT
calculations.\cite{Rondinelli:Nn,Duan:Prl,Burton:Prb,Burton:Prl,Stengel:Prl,Bristowe:cm}
The long-range Coulomb interaction is included via a layer-dependent potential $V(z)$,\cite{Stengel:Prl}
and within each layer the potential is assumed to be uniform for simplicity. This electrostatic potential
is determined via the Poisson equation.\cite{Brey:Prb,Yu:Prb09,Calderon:Prb08,Nanda:Prb,Calderon:Prb11}
In particular, the electric field between the $z$-th and ($z+1$)-th layers is determined by the net
charge [$Q+\sum_l^{1\leq l\leq z}(-n(l)+n_b)]$ counted from the FE interface, where $n(l)$ is
the $e_{\rm g}$ electronic density corresponding to the $l$-th layer,
and $n_b$ is the background (positive) charge density. Thus, the electrostatic potential
(with respect to the negative charge of electrons)
of each layer can be calculated via the relation:
\begin{equation}
V(z+1)=V(z)+\alpha[Q+\sum_l^{1\leq l\leq z}(-n(l)+n_b)],
\end{equation}
where $\alpha$ is the Coulomb coefficient which is inversely proportion
to the dielectric constant $\varepsilon$ [$\alpha=c/(\varepsilon t_0$),
where $c$ is the lattice constant, $\varepsilon$ is the dielectric constant, and $t_0$ is in unit of eVs as explained before].
In the following, $n_b$ is fixed at the value $0.7$ since typical manganites
are FM metals at this doping value, e.g. LSMO
and La$_{0.7}$Ca$_{0.3}$MnO$_3$ (LCMO).\cite{Tokura:Rpp}

In our computational study, the $12$-th layer is assumed to be sufficiently far
from the interface such that $V(z=12)$ is
set to be $0$ as the reference point of the electrostatic potential.
This choice, combined with a fixed chemical potential,
restores the system to its original bulk state for layers
far from the interface. A FM $t_{\rm 2g}$ background
is adopted to simulate the metallic channel in the FE-FET device.
The DE Hamiltonian (including the term with $V_in_i$) is diagonalized
to obtain the charge distribution $n(z)$, which is iterated
together with $V(z)$ until a self-consistent solution is reached.
After convergence in $n(z)$ and $V(z)$, the total grand potential (per u.c.) can be calculated as:
\begin{eqnarray}
\nonumber \Omega&=&\Omega_f-\frac{1}{2L_z}\sum_z^{1\leq z\leq L_z}V(z)n(z)-\frac{1}{2L_z}n_b\sum_z^{1\leq z\leq L_z}V(z)\\
&&-\frac{1}{2L_z}V(0)Q+\frac{J_{\rm AF}}{L_xL_yL_z}\sum_{<i,j>}\vec{S}_i\cdot\vec{S}_j,
\label{E.energy}
\end{eqnarray}
where $\Omega_f$ is the fermionic grand potential (per site), calculated from the diagonalization
eigenvalues.
The second term considers the reduction of the electrostatic Coulomb energy of the $e_{\rm g}$ electrons,
since it is doubly-counted in the first term. The third and fourth terms are the electrostatic Coulombic energies
of the positive background charge ($n_b$) and the FE surface charge, respectively. The last term describes the AFM SE energy, namely
the Heisenberg interaction among the localized spins.
A finite but low temperature $T$=$0.005t_0$ ($\sim30$ K) is used for the Fermi-Dirac distribution function smearing.

\section{Results and Discussion}
\subsection{Charge accumulation/depletion}
To investigate the screening effects in the FE-FET heterostructure, the results for
four values of $\alpha$ ($0.5$, $1$, $2$, and $4$) were compared. For each $\alpha$, the surface charge
$Q$ is initially set to zero to find the chemical potential where the average $e_{\rm g}$ density equals $n_b$.
With this chemical potential, $Q$ is then varied from $+0.4$ to $-0.4$ (in units of the elementary charge per cell).
Ideally, $|Q|$=$0.4$ corresponds to a FE polarization as large as $40$ $\mu$C/cm$^2$ (if the pseudocubic lattice
constant $c$ is set as $4$ \AA{}), which is a typical and reasonable value for standard FE oxide materials.

\begin{figure}
\centering
\includegraphics[width=0.4\textwidth]{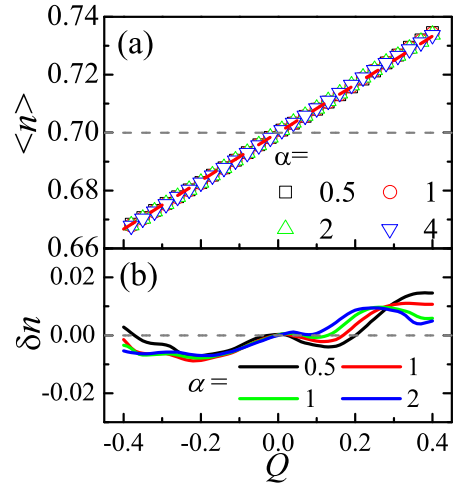}
\caption{(Color online) (a) The average $e_{\rm g}$ density $\langle n \rangle$ \textit{vs} $Q$. The (red) dashed line corresponds
to the fully screened case where $\langle n \rangle$=$n_b+Q/L_z$. Here only the electronic screening is considered, while
the ionic screening \cite{Chisholm:Prl} is neglected since its effect can be partially expressed by the dielectric constant, that
enters in $\alpha$,
and an effective $Q$. (b) The deviations of the $e_{\rm g}$ density from the fully screened limit,
where $\delta n=(\langle n_i \rangle -n_b)\times L_z-Q$. The maximum deviation ($|\delta n|$) is $<0.015$
for $\alpha$=$0.5$, which decreases to $<0.01$ for $\alpha$=$2$ and $\alpha$=$4$.}
\label{density}
\end{figure}

The screening effects correspond to the accumulation/depletion of charges near the interface. Under a positive (negative)
$Q$, more $e_{\rm g}$ electrons will be attracted to (repelled from) the interface. Since the chemical potential is fixed
in our simulation, the screening effect can also be obtained from the average $e_{\rm g}$ density as a function of $Q$,
as shown in Fig.~\ref{density}. This screening effect increases when $\alpha$ is increased,
which is concomitant with a stronger
electrostatic Coulomb interaction near the interface.
In the rest of the manuscript, $\alpha$=$2$ will be here adopted:
using $t_0$=$0.5$ eV and $c$=$4$ \AA{}, this $\alpha$ value corresponds to a relative permittivity $\varepsilon_r\approx45$,
which is quite reasonable to represent real materials.
Also note that $\alpha$=$2$ is already very close to the
fully screened case according to the results shown before.
It should be remarked that the total charge for the whole system is zero (i.e. the
combined FE gate and manganite channel are neutral)
although the gate and channel themselves are charge polarized.

\begin{figure}
\centering
\includegraphics[width=0.5\textwidth]{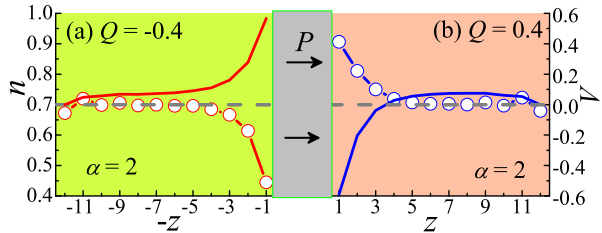}
\caption{(Color online) The $e_{\rm g}$ density profiles $n(z)$ (dots) and the electrostatic potential $V(z)$ (lines without dots).
The cases $Q$=$-0.4$ (left) and $Q$=$0.4$ (right) are shown together for better comparison.
The FE gate is in the middle and its polarization points to the right as indicated.
The original $e_{\rm g}$ density ($n_b=0.7$) is
shown  as dashed lines for better reference.}
\label{QVn}
\end{figure}

The screening effect is better observed by studying the $e_{\rm g}$ electron density profiles and their corresponding
electrostatic potentials in Fig.~\ref{QVn}. The $Q$=$+0.4$ and $-0.4$ cases are shown together for better comparison.
When $Q$=$+0.4$, then $V(z)$ becomes deep enough near the interface to accumulate considerably
more $e_{\rm g}$ electrons than in the bulk. In contrast,  when $Q$=$-0.4$, then
$V(z)$ is large and positive near
the interface, thus repelling those $e_{\rm g}$ electrons.
With $\alpha$=$2$, the screening of $e_{\rm g}$ electrons
is the most significant within a thin region near the interface,
typically involving just $2\sim3$ layers for the 2O DE model
employed here.

\subsection{Interfacial phase transitions}
\begin{figure*}
\centering
\includegraphics[width=0.9\textwidth]{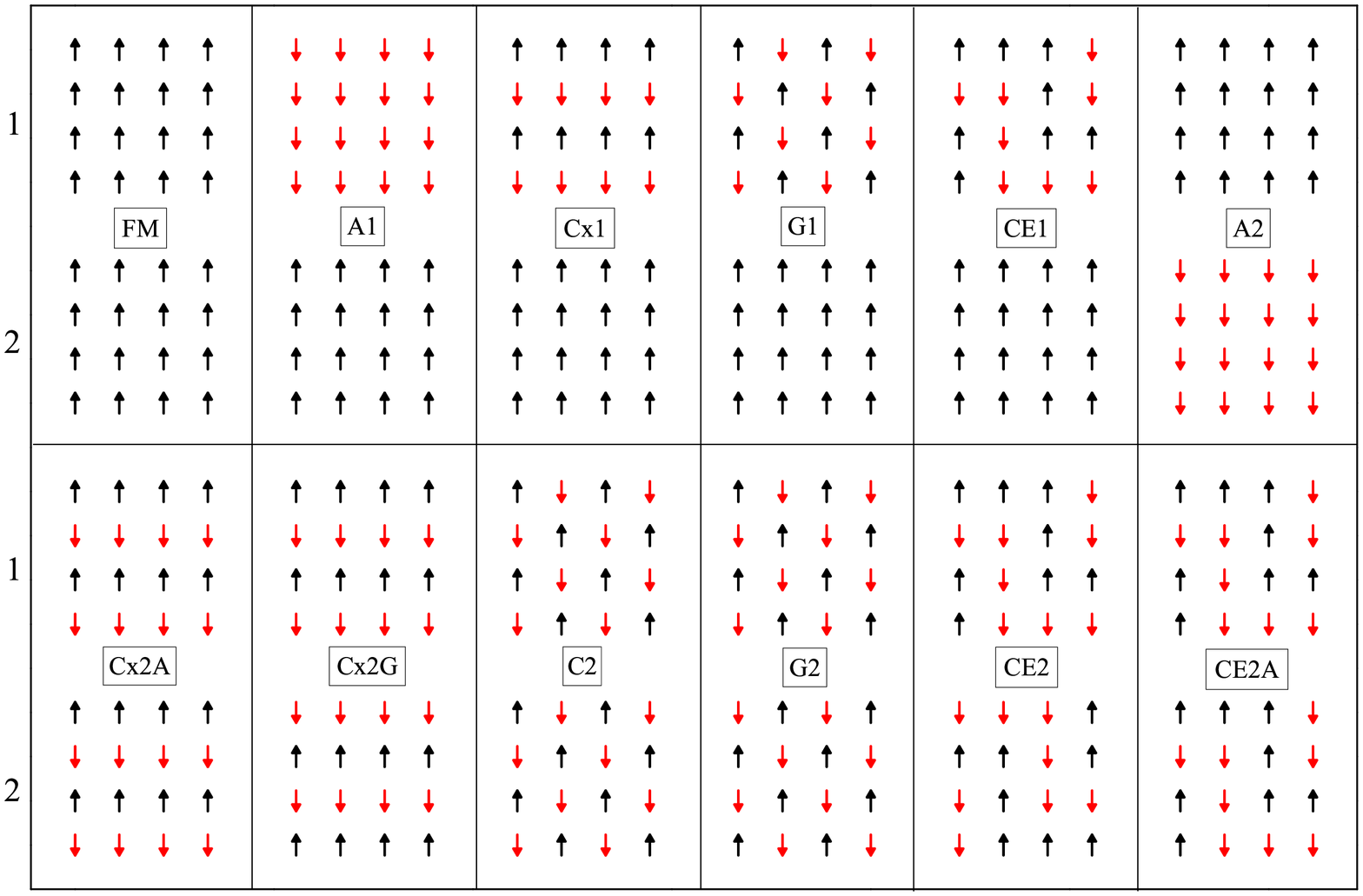}
\caption{(Color online) The candidates for the spin order at the two interfacial layers, as described
in the main text.
The layer indexes (1 and 2) counting from
the interface are shown on the left side of the figure.
The FM spin order is the original one, in the absence
of the surface charge $Q$.
In the rest of the panels, the spins pointing down are shown in red.
All spins in other layers ($z>2$) point ``up" for these variational states.
Here the choices for the spin candidate states
are not arbitrary but have clear correspondences to states already known to exist
in the bulk phase diagrams. In addition, some combinations of different
magnetic orders in the two layers
have also been included since the interfacial region may be different from the bulk.}
\label{spin}
\end{figure*}

Since the previous results show that the interfacial electronic density
can be substantially modulated by the FE polarization, then it is natural
to expect local phase transitions. The reason is that the phase diagrams of oxides are
usually highly sensitive to charge density variations,\cite{Dagotto:Sci}
i.e. density-driven phase transitions are well known to occur in bulk materials
when chemically doped to modify the electronic density.\cite{Burton:Prb,Burton:Prl}
To explore these possible phase transitions, the zero-temperature variational method
is here employed by comparing the total ground-state energy (Eq.~\ref{E.energy}) for
a variety of spin patterns. From Fig.~\ref{QVn}, it is clear that most of the
charge accumulation/depletion occurs within the first two layers near the FE interface.
Hence, for simplicity the several non-FM (collinear) spin patterns explored here
will only be proposed to exist in these two layers in our present variational calculation,
while the spins in the other layers remain fixed to be FM.
The candidate spin patterns in the two interfacial layers are shown in Fig.~\ref{spin}.

\begin{figure}
\centering
\includegraphics[width=0.4\textwidth]{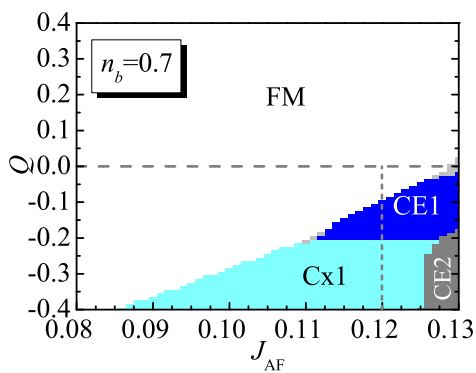}
\caption{(Color online) Ground state phase diagram for the interfacial layers in FE-FET,
obtained by the variational procedure described in the text. }
\label{pfet}
\end{figure}

The ground state phase diagram obtained in our calculations for the interfacial layers in FE-FET
is shown in Fig.~\ref{pfet}. According to this phase diagram, the original FM metallic
phase at $Q$=$0$ is stable when $J_{\rm AF}<0.128$,
while the boundary between the FM and A-type AFM phases is at $J_{\rm AF}$=$0.13$ for the calculation
representing the bulk (see Fig.~\ref{bulk} later in this publication).
These two almost identical values suggest that the lattice size effects and surface effects
are negligible in our simulation of FE-FET.

By adjusting the FE polarization (i.e. by modifying the surface charge $Q$) in the FE-FET setup,
in the present variational effort it has been observed that the interfacial spins have a transition
to arrangements different from the original FM state. This is
the main result of our publication. For example, the CE1 and Cx1 orders are stabilized and replace
the FM state
in sequence with increasing negative $Q$ when $J_{\rm AF}\sim0.12t_0$, as shown
in Fig.~\ref{pfet}. In contrast, the FM order remains robust under a positive $Q$, thus establishing
an asymmetry in the response of the system to the FE polarization orientation
that is of value for applications.

\begin{figure}
\centering
\includegraphics[width=0.4\textwidth]{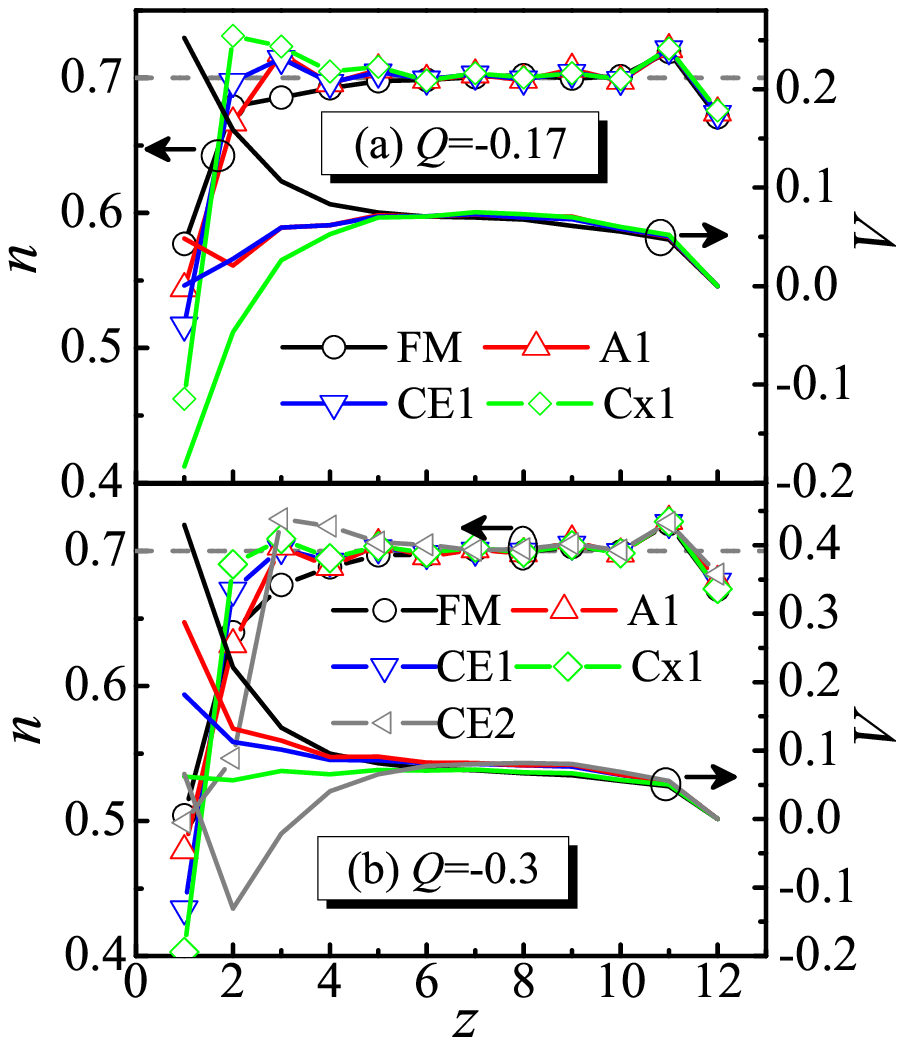}
\caption{(Color online) The $e_{\rm g}$ density profiles (left axes)
and the corresponding electrostatic potentials (right axes) of the 2O model studied here.
Panel (a) is for $Q$=$-0.17$ while panel (b) is for $Q$=$-0.3$.
The ground states (CE1 state in (a) and Cx1 state in (b)) provide
the best screening effect, i.e. a smooth potential $V(z)$ varying $z$.}
\label{pp}
\end{figure}

The FE screening effect plays an important role to determine the dominant
interfacial spin order, that is competing with the DE mechanism that favor ferromagnetism.
Considering $J_{\rm AF}$=$0.12$ as example, when $Q<0$ the CE1 and Cx1 orders
can accommodate more holes near the interface than the original FM state,
thus reducing the Coulomb potential pronouncedly, as shown in Fig.~\ref{pp}.
In simple words, the system chooses
an interfacial state which can screen the FE polarization rather well.

\subsection{Comparison with bulk properties}
\begin{figure}
\centering
\includegraphics[width=0.4\textwidth]{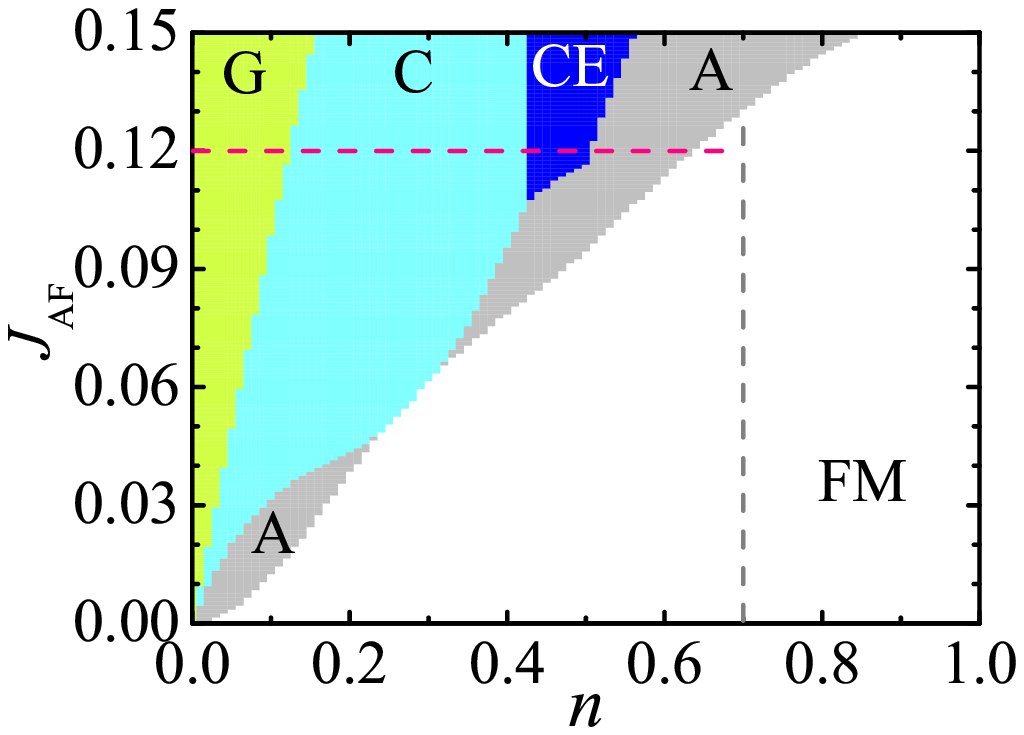}
\caption{(Color online) The ground state phase diagram of the 2O DE model for manganites
in the bulk, which is calculated using the variational method described in the text.
All DE energies are obtained from analytical band structures. The phase boundary between
the FM and A states at $n$=$0.7$ is illustrated by the vertical dashed line,
while the horizontal dashed line shows the phase transition at $J_{\rm AF}$=$0.12$
obtained by changing the $e_{\rm g}$ electron density. A, C, CE, and G denote the
typical AFM phases found in manganites.\cite{Tokura:Rpp}}
\label{bulk}
\end{figure}

For comparison, the ground state of the bulk is also calculated
using the standard 2O DE model, under a similar variational approximation with states now covering the
whole system. This information can be used as a guide to explore the interfacial spin orders that may
be of relevance in the FE-FET setup. The results are shown in Fig.~\ref{bulk}.
Considering the simplicity of the model (with only two competing
NN interactions: DE \textit{vs} SE), this phase diagram agrees fairly well
with the experimental perovskite manganite results.\cite{Tokura:Rpp}
The most typical phases found in bulk manganites, namely the FM
and various AFM states (A-, C-, G-, and CE types),
appear in the proper $e_{\rm g}$ density and bandwidth regions, providing support to the qualitative
accuracy of our calculations.

\begin{figure}
\centering
\includegraphics[width=0.5\textwidth]{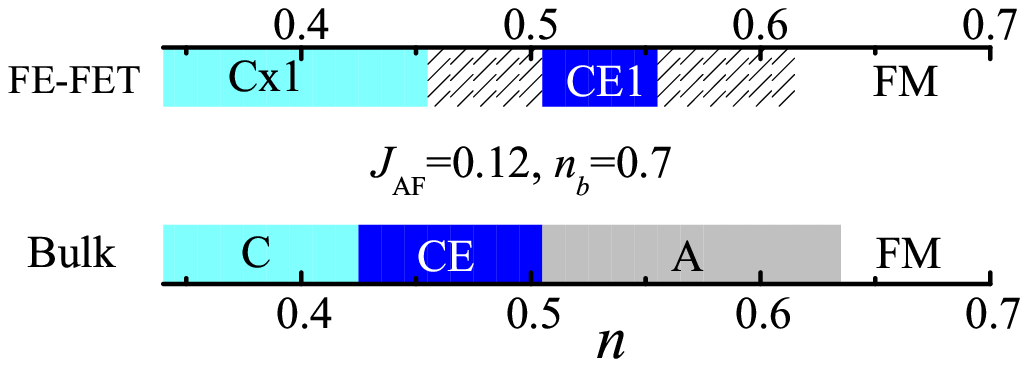}
\caption{(Color online) Comparison of the transitions found in the FE-FET heterostructure
and in the bulk, at $J_{\rm AF}=0.12$. According to the bulk phase diagram,
with decreasing electronic density the bulk system turns from the FM phase
into an A-type AFM state, followed by a CE phase and then by a C-type AFM state.
However, in the FE-FET setup, from the original FM phase and with increasing $|Q|$,
the spins in the first interfacial layer directly  jump to the CE order, and then to the
Cx1 order. Moreover, the shaded regions were found to be unstable
due to phase separation tendencies in the FE-FET case.}
\label{phase}
\end{figure}

Considering $J_{\rm AF}$=$0.12$ as an example, Fig.~\ref{phase} compares
the spin order transitions in the bulk and in the FE-FET. In the bulk's phase
diagram, by reducing the $e_{\rm g}$ density from $n$=$0.7$, the system transitions
from a FM phase  to an A-type AFM state at $n$=$0.63$, then from A to CE at $n$=$0.5$, and
from CE to C-type AFM one at $n$=$0.42$. In the FE heterostructure, on the other hand,
the system changes from FM to CE1 at $Q$=$-0.1$ ($n$=$0.62$ in FM and $n$=$0.55$ in CE1), and
then from CE1 to Cx1 at $Q$=$-0.19$ ($n$=$0.51$ in CE1 and $n$=$0.45$ in Cx1).
There are several interesting aspects in this interfacial phase transitions.
First, the ``critical'' $e_{\rm g}$ densities are found to be {\it different} between
the bulk and the FE-FET heterostructure. Second, the fragile A-type AFM state is absent
in the heterostructure geometry. Third, in the heterostructure the interfacial electronic
density jumps at the locations of the spin order transitions, causing some density regions
to be unreachable (i.e. they are unstable). Such density discontinuities originate
from the well-known electronic phase separation tendencies
in manganites,\cite{Dagotto:Prp,Dagotto:Bok,Dagotto:Njp} a phenomenon that
does not have an analog in semiconducting devices. Last but not least, the CE1 and Cx1
states predicted here have not been considered in previous DFT studies, since these states
typically need larger in-plane cells than previously analyzed with DFT.
These two interfacial states, CE1 and Cx1, may exist particularly
in those manganite channels with
relative narrow bandwidths, such as LCMO.

There are two main reasons for the differences observed here in the phase diagrams between
the bulk and the heterostructures. The first reason is the FE screening effect,
as shown in Fig.~\ref{pp}, namely the ground state near the interface is determined
not only by the competition between the DE kinetic energy and the SE energy as in
the bulk, but also by the electrostatic potential energy. Second, since the spin order
transitions occur only near the interface, the global phases shown here, except
for the FM one, are actually ``artificial" phase separated states involving
a combination of the bulk and the interfacial states, combination that may be more
stable than the homogeneous spin orders in the FE-FETs. Thus, these examples show
that it is not enough to simply guess the interfacial spin orders from those in
the bulk phase diagrams with only homogeneous phases: new states may emerge at the interfaces.

Furthermore, it should be noted that these phase transitions may be
even more complex than our calculations suggest. For instance, other states
beyond the candidates considered here, for instance involving canted spins
and thicker interfacial layers, may become stable in some regions. To reveal
additional details of these interfacial phase transitions, unbiased (and very CPU time
consuming) studies involving Monte Carlo simulations should be performed in the future,
including electron-phonon couplings and finite temperature effects. However, the results
discussed here are already sufficient to clearly show that the original FM phase
is indeed unstable toward other phases at the interface with a FE,
which was the main goal of this publication.

\subsection{Spin flip \textit{vs} spin rotation}

Although the studies described above already clearly show that
interfacial phase transitions away from the FM state will occur
by tuning the FE polarization, the fine details of these phase transitions
remain unclear. Do these spins flip abruptly from one configuration to the other
or do they rotate gradually
upon increasing $|Q|$? Are there any canted spin states tendencies
besides the collinear spin candidates considered here?
Reaching a full answer to these questions is computationally very difficult
at the current state of typical Monte Carlo simulations with an effort that grows like
the fourth power of the number of sites $N$. However, some studies
concerning spin rotation {\it vs.} spin flip tendencies
can still be carried out in a variational manner, as described below.

\begin{figure}
\centering
\includegraphics[width=0.4\textwidth]{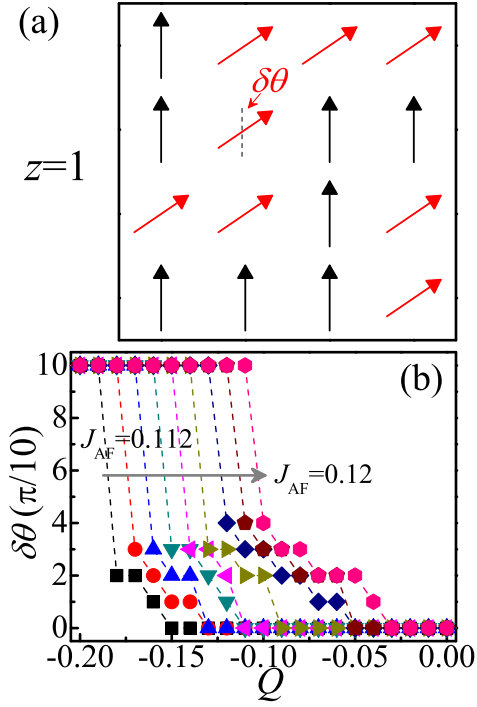}
\caption{(Color online) Possible spin flip/rotation process from the FM state ($\delta\theta=0$)
to the CE1 state ($\delta\theta=\pi$) during the interfacial phase transitions discussed before.
(a) Sketch of the angle $\delta\theta$ used in the calculation.
(b) Results for the nine sets of $J_{\rm AF}$ values ($0.112-0.12$) considered here.}
\label{Q}
\end{figure}

As shown in Fig.~\ref{pfet}, the FM order turns into the CE1 order
upon increasing $|Q|$, which involves only one interfacial layers.
During this transition, half of the spins in the first layer flip to
``down'' spins in the final CE1 state.
For simplicity, let us assume that this phase transition (spin flip) occurs via
an in-plane spin rotation. To reach the CE1 order, the spins in the first layers
are partitioned into CE type zigzag chains. Half of those zigzag chains are assumed to
rotate synchronously, namely they are characterized by an unique
spin angle $\delta\theta$. Using the variational method, $\delta\theta$ can be determined
as a function of $Q$, as shown in Fig.~\ref{Q}. Since the phase boundary between the
FM and CE1 states also depends on the SE coupling ($J_{\rm AF}$),
the spin flip/rotation process varies with $J_{\rm AF}$.
For all $9$ sets of $J_{\rm AF}$ shown here, the ``speeds'' of the spin rotations
are not uniform. Note that a sharp jump of $\delta\theta$ always exists in each of the
curves. There are only a few spin canted states that are stable as intermediate states
during the spin rotation process, most of which exist near the FM side
(i.e. $\delta\theta\sim0$). Thus, the spin canting process does not seem
to be very robust, at least according to our qualitative calculations.
Instead, a sudden spin flip may be the preferred process for the interfacial phase transitions.

A better characterization of the spin flip \textit{vs}. spin rotation tendencies relates
with the first-order vs. second-order transition character of the process.
From Fig.~\ref{Q}, it seems that both spin flip and spin rotation are allowed.
However, the canting angles $\delta\theta$'s are restricted near $0$ in the spin flip case.
Thus, there seems to occur a first-order transition between a FM-like state
(with $\delta\theta\sim0$) and an AFM state (with $\delta\theta\sim\pi$).
Of course, more powerful unbiased computational methods should be used
to confirm this conclusion.

\subsection{Resistive switching}
\begin{figure}
\centering
\includegraphics[width=0.4\textwidth]{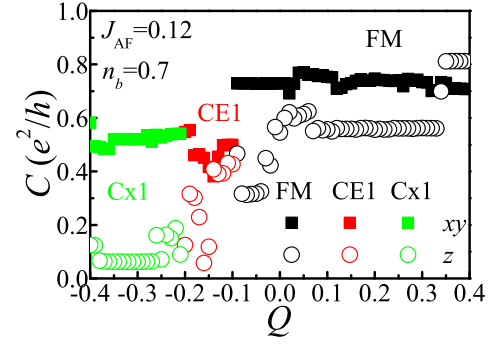}
\caption{(Color online) The Kubo conductance as a function of $Q$.\cite{Verges:Cpc}
Both the in-plane ($xy$) and out-of-plane ($z$) conductances are shown.}
\label{cc}
\end{figure}

Although the charge accumulation/depletion and associated local phase transitions induced
by the switch of the FE polarization orientation
occur only near the interface, these transitions lead to a global change in the conductance
of the metallic channel (e.g. LSMO) when the FE polarization is flipped.
This resistive switch effect should be bipolar, due to the asymmetric phase diagram found
in our calculations,
as shown in Fig.~\ref{pfet}. This effect should also be anisotropic, because
in a metallic channel, when the interfacial layers become less conducting due
to the previously described phase transitions, the out-of-plane conductance
will be seriously suppressed basically
due to the spin valve effect \cite{Burton:Prl,Bristowe:cm} while  the in-plane
transport will be only weakly affected, as shown in Fig.~\ref{cc}. It should be
noted that here a good FM metallic channel is used, while more prominent resistive
changes are expected to occur in those systems which are close to metal-insulator phase boundaries.

Besides the changes of the resistivity, magnetoelectric effects
have also been observed in experimentally studied
FE-FET heterostructures.\cite{Molegraaf:Am,Vaz:Prl,Vaz:Apl}
Qualitatively, the change of the magnetization can be understood
via the local phase transitions
near the interface when the FE polarization is flipped (Fig.~\ref{pp} and Fig.~\ref{Q}), as described
in this manuscript.

Finally, it should also be noted that our current effort provides just a starting point
to study the FE field-effect heterostructures
 with the use of  model Hamiltonians. Additional realistic effects in real heterostructures
were neglected in the present work, such as lattice structural distortions and
chemical bonding effects. Thus, the current predictions may be not as accurate
as those reached with DFT calculations for some particular materials.
The main relevance of the present model-based study is that it can provide
overall tendencies for a material family. The study of the effect of more
realistic interactions and the inclusion of finite-temperature effects
can be achieved in future calculations based on the model described here.

\section{Conclusions}

In summary, a microscopic model Hamiltonian for the FE oxide~-~FM metallic oxide heterostructures,
a prototypical FE-FET system, has been here studied. The FE field effect is modeled via
the electrostatic Coulomb potential in the FM oxide. Using a self-consistent calculation
and the variational method, an interfacial charge accumulation/depletion is found by tuning
the magnitude and sign of the FE polarization. Phase transitions at the interface
have been observed here by modulating the electronic charge density of the metallic component
by varying the FE polarization. Our present effort provides a starting point to study
the FE field effect via model Hamiltonians. Our results clearly present some common
similarities with previous DFT effort, confirming their main results.  However,
the framework is conceptually different and the results reported here are
not identical to those of DFT. Moreover, our model is generic and it can be adapted
to study a variety of other oxide heterostructures involving ferroelectrics, particularly
those where the metallic component has a strongly correlated electronic character.

\section{Acknowledgments}

We thank Ho Nyung Lee for helpful discussions. The
work of S.D. was supported by the 973 Projects of China (2011CB922101,
2009CB623303), NSFC (11004027), and NCET (10-0325). R.Y. was supported by
the NSF grant (DMR-1006985) and the Robert A. Welch Foundation (C-1411).
X.Z. and E.D. were supported by the U.S. Department of Energy,
Office of Basic Energy Sciences, Materials Science and Engineering Division.

\bibliographystyle{apsrev4-1}
\bibliography{../ref}

\end{document}